\documentclass{roajarticle}
\usepackage{graphicx}
\usepackage{mynatbib}
\usepackage{upgreek}
\usepackage{upgreek2}
\usepackage{amsmath}
\usepackage{subcaption}
\usepackage[section]{placeins}
\newcommand{\volume}{\textbf{31}}
\newcommand{\numarul}{1}
\newcommand{\volyear}{2021}
\graphicspath{ {./figs/} }

\title{Study of the relation between luni-solar periodicities and earthquake events}

\author[1]{Ibnu  \MakeTextUppercase{Nurul Huda}}
\author[2]{Jean  \MakeTextUppercase{Souchay}}
\affil[1]{Department of Astronomy and Bosscha Observatory, FMIPA, Institut Teknologi Bandung, Jalan Ganesha 10, Bandung 40132, Indonesia}
\affil[2] {SYRTE, Observatoire de Paris, PSL Research University, CNRS UMR 8630, Paris, France}
\affil[ ] {Email: ibnu@as.itb.ac.id}

\keywords{Earthquake interaction -- Tides and planetary waves -- Statistical test}

\begin{document}
\maketitle

\begin{abstract}
We revisit the work of \citet{kilstonknopoff1983} to study the correlation of earthquakes with the main lunisolar tidal components in the limited zone of the Southern California region, with a considerably bigger amount of data. By adopting the same methodology as these authors, we confirm their results, i.e. that strong earthquakes with M $\geq 6$ show a statistically significant correlation with half day and precession cycles, with respective periods 12 hr. and 18.6 y., while weaker earthquakes show a correlation with the annual cycle. Moreover we extended the study to the analysis of bi-modal phase to phase diagrams, leading to a very high probability of a lack of earthquake occurrences when both phases of the 12 hr. and 18.6 y. cycles are inside the [0 - 0.5] interval. Furthermore, we demonstrate that the statistical method is not really appropriate to study the correlation with the semi-lunar day  and the fortnightly tidal components.
\end{abstract}

\section{Introduction}

The relation between lunisolar cycle and earthquake occurrences has been a long-standing debate for more than one century. Indeed the Earth tide produced by the Moon and the Sun is subject to very well modelized oscillations and the corresponding tidal stress is superimposed on tectonic stress. Although the amplitude of the tidal stress changes is generally considerably smaller than the stress drop related to earthquakes, typically of the order of $10^3$ Pa., its frequency is generally much larger. Therefore when the stress level is critical with respect to the release of an earthquake, the rapid additional stress change due to the Earth tide may trigger an earthquake. 

Based on this principle, there have been numerous searches for the relation between Earth tides and earthquake occurrences. Some authors concluded that there is a correlation between these two phenomena \citep[see, e.g.,][]{ryalletal1968,tamrazyan1974,youngzurn1978,nikolaev1995,saltykovetal2004} whereas others reject this conclusion  \citep[see, e.g.,][]{knopoff1964,shlien1972,shudde1977}. In general, the correlation of earthquakes with lunisolar periodicities is studied in two ways. First, several authors have devoted their efforts to the correlation on a global scale \citep[see, e.g.,][]{tanakaetal2002,metivieretal2009}. Meanwhile, since the nature of the correlations seems to be strongly dependent on the faulting mechanism involved, other authors conducted the study by considering a limited zone rather than a global one. Such local investigations were done for instance for the Southern California area \citep{kilstonknopoff1983}, the Italy territory \citep{Palumbo1986}, the Kamchatka area \citep{shirokov1983}, the Vrancea area in Romania \citep{souchaystavinschi1997,stavinschisouchay2003,cadicheanuetal2007}, Taiwan \citep{chenetal2012}, Hellenic Arc  \citep{vergosetal2015}, and Colombia \citep{moncayoetal2019}.

Here we revisit the work of \citet{kilstonknopoff1983}, hereafter named as K \& K, to study the earthquake correlation with lunisolar cycles in the limited zone of the Southern California. Following this last paper, the reason to search such a correlation stays in geotectonic considerations: as it is the case for the San Andreas fault, the majority of faults in this region are aligned along an NW-SE direction, and their characteristics are right-lateral strike-slip. As a result, extensional tidal stresses in the east-west direction increase the shear stress at the origin of the faulting. Note that Earth tides are the largest known sources of oscillatory stresses with relatively short periods, as it is asserted by K \& K. Since east-west extensional stress is maximized for the Moon and the Sun respectively at the Moonrise/set and the Sunrise/set we can consider the intuitive hypothesis that earthquakes occur preferentially during these instants, as those corresponding to maxima of other leading tidal components. It should be all the more the case when both bodies, Moon and Sun are close to the configuration of syzygies, that is to say near the new Moon and the full Moon, when their gravitational action is combined. By using the same statistical approach as K \& K, we take advantage of the better quality of the seismologic data available as well as new data gathered during roughly four more decades after K \& K publication. 

This paper is constructed as follows. In Section \ref{data_method}, we explain in detail both the contents of the seismologic data used and the statistical method based on $\chi^2$ analysis. The phase distribution of the earthquake events relative to the main lunisolar tidal cycles is analyzed in Section \ref{phase}. Section \ref{correlation} presents the estimation of correlations between Earth tides components and earthquake occurrences through the same statistical test as used by K \& K. Section \ref{sensitivity} discusses the sensitivity of the results with respect to the period of the cycle considered. Finally, the conclusion is given in Section \ref{conclusions}.

%%%%%%%%%%%%%%%%%%%%%%%%%%%%%%%%%%%%%
\section{Data and method} \label{data_method}
%%%%%%%%%%%%%%%%%%%%%%%%%%%%%%%%%%%%%

We use the earthquake data from United States Geological Survey (\url{https://earthquake.usgs.gov/}), by using, in particular, the catalogues CI \citep{southern2013southern} and USHIS \citep{StoverCoffman1993}, with the data starting from 1800 to 2020. We limit our investigations to the region inside the latitude interval from $33.0^{\circ}$ to $36.0^{\circ}$  and the longitude interval westward from $-115.5^{\circ}$ to $-120.3^{\circ}$. To minimize the possible contamination of our data set by aftershocks, we limit the study by considering only the earthquakes with magnitude M $\geq 5.3$ and eliminate earthquake aftershocks that occur close to strong earthquakes M $\geq 6$. Thus by using this restriction, we get 68 isolated earthquake events. 32 of them have a magnitude M $\geq 6$ and the remaining 36 ones range in the interval $5.3 \leq M < 6$. These numbers are almost twice bigger than in the data used by K \& K. Tables \ref{tab:earthquake_list_6} and \ref{tab:earthquake_list_5} display respectively the earthquake events found in both magnitude intervals.

\begin{table}
\small
\caption{List of earthquake events with M $\geq 6$. The symbol * signifies that the event was taken into account by K \& K.}
\begin{center}
{\renewcommand{\arraystretch}{0.7}

\begin{tabular}{lcccc}
\hline
\hline
Date & hr:min & lat ($^{\circ}$) & long ($^{\circ}$) & mag \\
& (UTC) & & & \\
\hline
2019-07-06    &  03:19  &  35.8 & -117.6 & 7.1 \\     
2010-07-07    &  23:53  &  33.4 & -116.5 & 6.0 \\
1999-10-16    &  09:46  &  34.6 & -116.3 & 6.6 \\
1994-01-17    &  12:30  &  34.2 & -118.5 & 6.4 \\
1992-06-28    &  11:57  &  34.2 & -116.4 & 7.3 \\
1987-11-24    &  13:15  &  33.0 & -115.8 & 6.0 \\
1971-02-09 *  &  14:00  &  34.4 & -118.4 & 6.4 \\
1968-04-09 *  &  02:29  &  33.2 & -116.1 & 6.4 \\
1954-03-19 *  &  09:54  &  33.3 & -116.2 & 6.2 \\
1952-07-21 *  &  11:52  &  35.0 & -119.0 & 7.2 \\
1948-12-04 *  &  23:43  &  33.9 & -116.4 & 6.5 \\
1947-04-10 *  &  15:58  &  35.0 & -116.6 & 6.2 \\
1946-03-15 *  &  13:49  &  35.7 & -118.1 & 6.3 \\
1937-03-25 *  &  16:49  &  33.4 & -116.3 & 6.0 \\
1933-03-11 *  &  01:54  &  33.6 & -118.0 & 6.3 \\
1925-06-29 *  &  14:42  &  34.3 & -119.8 & 6.3 \\
1923-07-23 *  &  07:30  &  34.0 & -117.3 & 6.3 \\
1922-03-10    &  11:21  &  35.8 & -120.3 & 6.5 \\
1918-04-21 *  &  22:32  &  33.8 & -117.0 & 6.8 \\
1916-11-10    &  09:11  &  35.5 & -116.0 & 6.1 \\
1910-05-15    &  15:47  &  33.7 & -117.4 & 6.0 \\
1908-11-04    &  08:37  &  36.0 & -117.0 & 6.5 \\
1899-12-25    &  12:25  &  33.8 & -117.0 & 6.4 \\
1899-07-22    &  20:32  &  34.3 & -117.5 & 6.5 \\
1892-05-28    &  11:15  &  33.2 & -116.2 & 6.3 \\
1890-02-09    &  12:06  &  33.4 & -116.3 & 6.3 \\
1883-09-05    &  12:30  &  34.2 & -119.9 & 6.0 \\
1858-12-16    &  10:00  &  34.0 & -117.5 & 6.0 \\
1857-01-09 *  &  16:24  &  35.7 & -120.3 & 7.6 \\
1855-07-11    &  04:15  &  34.1 & -118.1 & 6.0 \\
1812-12-21    &  19:00  &  34.2 & -119.9 & 7.1 \\
1800-11-22    &  21:30  &  33.0 & -117.3 & 6.5 \\
\hline
\hline
\end{tabular}}
\label{tab:earthquake_list_6}
\end{center}
\end{table}

\begin{table}
\small
\caption{List of earthquake events with  $5.3 \leq$ M $< 6.0$. The symbol * signifies that the event was taken into account by K \& K.}
\begin{center}
{\renewcommand{\arraystretch}{0.7}

\begin{tabular}{lcccc}
\hline
\hline
Date & hr:min & lat ($^{\circ}$) & long ($^{\circ}$) & mag \\
& (UTC) & & & \\
\hline
2018-04-05    &  19:29  &  33.8 & -119.7 & 5.4 \\
2012-08-26    &  20:57  &  33.0 & -115.5 & 5.5 \\
2008-07-29    &  18:42  &  33.9 & -117.8 & 5.9 \\
2005-06-12    &  15:41  &  33.5 & -116.6 & 5.6 \\
2004-09-28    &  17:15  &  35.8 & -120.4 & 5.7 \\
1995-09-20    &  23:27  &  35.8 & -117.6 & 5.8 \\
1995-08-17    &  22:39  &  35.8 & -117.7 & 5.4 \\
1991-06-28    &  14:43  &  34.3 & -118.0 & 5.9 \\
1990-02-28    &  23:43  &  34.1 & -117.7 & 5.5 \\
1988-06-10    &  23:06  &  34.9 & -118.7 & 5.4 \\
1987-10-01    &  14:42  &  34.1 & -118.1 & 5.9 \\
1986-07-08    &  09:20  &  34.0 & -116.6 & 5.6 \\
1981-09-04    &  15:50  &  33.7 & -119.1 & 5.3 \\
1981-04-26    &  12:09  &  33.1 & -115.6 & 5.6 \\
1980-02-25 *  &  10:47  &  33.5 & -116.5 & 5.5 \\
1973-02-21 *  &  14:45  &  34.1 & -119.0 & 5.9 \\
1970-09-12 *  &  14:30  &  34.3 & -117.5 & 5.4 \\
1969-04-28 *  &  23:20  &  33.3 & -116.3 & 5.8 \\
1961-01-28 *  &  08:12  &  35.8 & -118.0 & 5.3 \\
1955-12-17 *  &  06:07  &  33.0 & -115.5 & 5.4 \\
1954-01-12 *  &  23:33  &  35.0 & -119.0 & 5.9 \\
1950-07-28 *  &  17:50  &  33.1 & -115.6 & 5.4 \\
1949-05-02 *  &  11:25  &  34.0 & -115.7 & 5.9 \\
1947-07-24 *  &  22:10  &  34.0 & -116.5 & 5.5 \\
1946-07-18 *  &  14:27  &  34.5 & -116.0 & 5.6 \\
1946-01-08 *  &  18:54  &  33.0 & -115.8 & 5.4 \\
1945-08-15 *  &  17:56  &  33.2 & -116.1 & 5.7 \\
1945-04-01 *  &  23:43  &  34.0 & -120.0 & 5.4 \\
1944-06-12 *  &  11:16  &  34.0 & -116.7 & 5.3 \\
1943-12-22 *  &  15:50  &  34.3 & -115.8 & 5.5 \\
1943-08-29 *  &  03:45  &  34.3 & -117.0 & 5.5 \\
1941-11-14 *  &  08:41  &  33.8 & -118.3 & 5.4 \\
1941-07-01 *  &  07:50  &  34.4 & -119.6 & 5.9 \\
1940-05-18 *  &  05:03  &  34.1 & -116.3 & 5.4 \\
1938-05-31 *  &  08:34  &  33.7 & -117.5 & 5.5 \\
1933-10-02 *  &  09:10  &  33.8 & -118.1 & 5.4 \\
\hline
\hline
\end{tabular}}

\label{tab:earthquake_list_5}
\end{center}
\end{table}

To analyse the statistical relationship between lunisolar tidal cycles and earthquake events, we use the same $\chi^2$ algorithm as in the work of K \& K. It is based on the significance of a cyclic Poisson process with respect to a homogeneous Poisson process evaluated starting from the hypothesis of a cyclic behaviour with a known phase shift. It enables us to prove the presence of a clustering of events around a specific phase value with respect to a specific oscillation with a given period. Of course this method assumes that the triggering of earthquakes might occur preferentially in some part of a fundamental cycle characterizing the earth tides. If $N$ is the total number of events which are being considered, then the $\chi^2$ parameter is defined as :
\begin{equation}\label{eq:chi}
    \chi^2 = \frac{2}{N}\left(\sum^n_{i=1}\cos{\Phi_i}\right)^2.
\end{equation}
Here $\Phi_i = 2\pi t_i/T$ is the phase, where $t_i$ is the time of occurrence of the $i$th. earthquake relative to the time of maximum possible correlation and $T$ is the period of the oscillation. A zero phase characterizes an earthquake occurring at a peak of the cycle. A null hypothesis can be rejected if $\chi^2 > 3.84$. In other words, such a value of $\chi^2$  means that there is a correlation between the lunisolar cycle considered and the earthquakes at a 95\% confidence level. In the same manner, a value of $\chi^2 > 2.71$ validates a correlation at a 90\% confidence level. 

We focus our analysis into five lunisolar cycles. First, we studied the correlation with the average time of interval between rise and fall of the sun, hereafter named "semi-diurnal" period. Here the zero phase is taken at 6 a.m and 6 p.m local time and the oscillation period is 12 hrs. Second, we consider the cycle of the precession of the line of nodes of the Moon orbit with a 18.613 years period and corresponding by far to the most significant tidal oscillation at low frequency. Here the zero phase is taken on January 12th., 1932.  We also estimated the correlation with other lunisolar cycles such as the mean fortnightly component (period = 14.77 d.; zero phase = full/new moon), the mean semi-lunar day (period: 12.42 hrs.; zero phase = lunar rise/set), and the annual cycle (period = 1 y.; zero phase = December 21th.). 

%%%%%%%%%%%%%%%%%%%%%%%%%%%%%%%%%%%%%%%%%%%%%
\section{Phase analysis} \label{phase}
%%%%%%%%%%%%%%%%%%%%%%%%%%%%%%%%%%%%%%%%%%%%%
\begin{figure*}
\centering
\includegraphics[width=0.8\hsize]{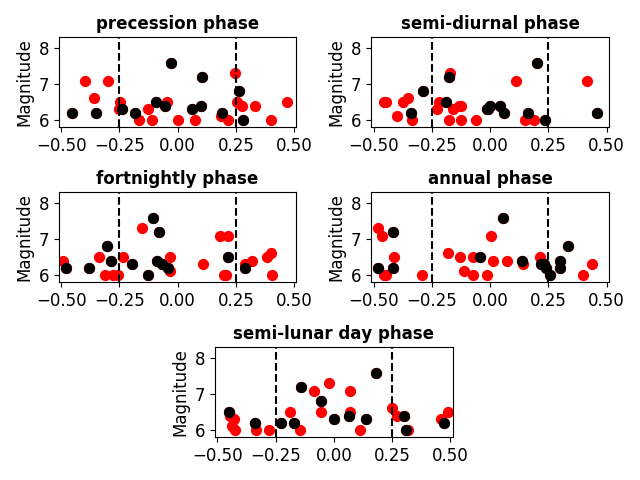}
\caption{Phase distribution of earthquakes occurrence ($\geq 6$ M) with respect to each lunisolar cycle. Zero phase corresponds to the maximum peak of the cycle. Here \textcolor{black}{$\bullet$} represents earthquake taken into account by K \& K.}
\label{fig:phasevsmag_all_strong}
\end{figure*}

\begin{figure*}
\centering
\includegraphics[width=0.8\hsize]{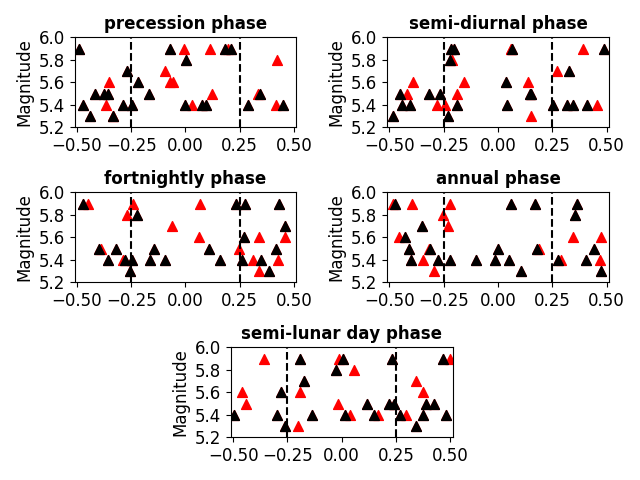}
\caption{Phase distribution of earthquake $5.3 \leq M < 6$ with respect to each lunisolar cycle. Zero phase corresponds to the maximum peak of the cycle. Here \textcolor{black}{$\blacktriangle$} represents the earthquakes taken into account by K \& K. }
\label{fig:phasevsmag_all_weak}
\end{figure*}

Phase distribution can be used as a rough approach to identify the clustering in the earthquake events. To calculate the phase $\Phi$, we use the relation $\Phi_i = 2\pi t_i/T$ and we re-scale the phase to the range (-0.5, 0.5). Then the phase distribution is classified into two groups: the first one corresponds to  $-0.25 \leq \Phi_i \leq 0.25$ while the second one is outside this range. 

Figures \ref{fig:phasevsmag_all_strong} and \ref{fig:phasevsmag_all_weak} show the phase distribution of the earthquake events relative to each lunisolar cycle considered respectively for the earthquakes with amplitude M $\geq 6$ and for those with $5.3 \leq M < 6$. For the precession cycle, we remark that strong earthquakes are concentrated in the central half of the diagram with 18 events inside against 12 events outside while weak earthquakes are more evenly distributed. A similar inequality also appears in more prominent manner in the case of the semi-diurnal cycle, with 21 earthquakes inside the central half-zone against only 9 outside. Notice that clustering for both cycles has been recognized by K \& K even with a noticeably smaller number of events taken into account.  Indeed they found that from 13 large events recorded with $M \geq 6$, 10 of them were located inside the central zone for the precession cycle and 9 events inside in the case of semi-diurnal. Meanwhile, for other cycles, the events are more evenly distributed both for strong and weak earthquakes.  

\begin{figure}
	\centering
	\begin{subfigure}[b]{0.49\textwidth}
		\includegraphics[width=1\textwidth]{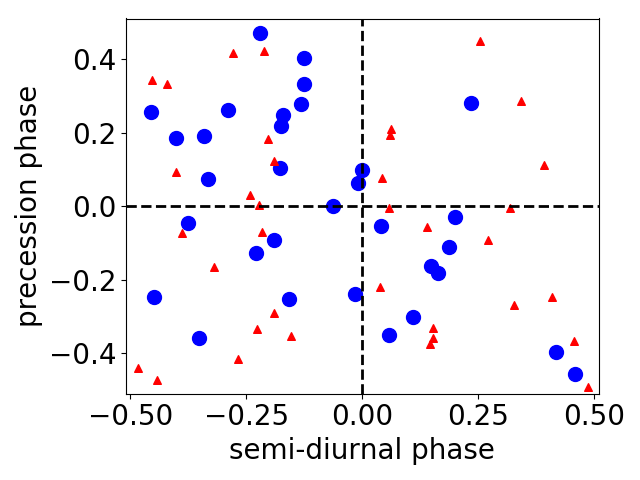}
		\caption{ }
		\label{fig:phasevsphase_1a}
	\end{subfigure}
	\begin{subfigure}[b]{0.49\textwidth}
		\includegraphics[width=1\textwidth]{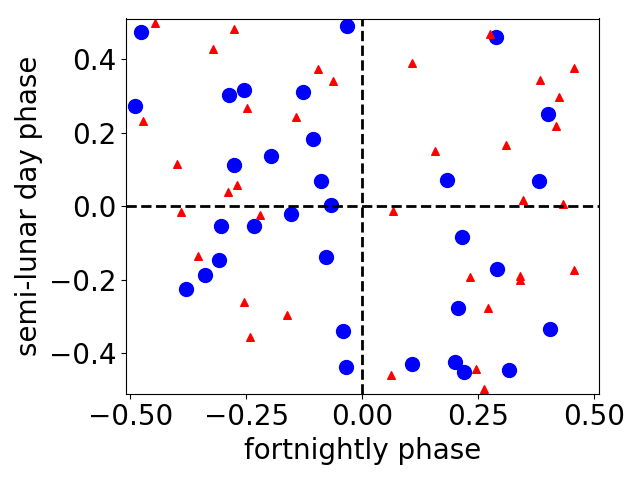}
		\caption{ }
		\label{fig:phasevsphase_1b}
	\end{subfigure}
\caption{Phase versus phase of two cycles. Here \textcolor{blue}{$\bullet$} corresponds to the earthquake $\geq 6$ M while the earthquake $< 6$ is represented by \textcolor{red}{$\blacktriangle$}.}
\label{fig:phasevsphase_1}
\end{figure}

Furthermore, we analyse the possibility of a link between two cyclic tidal components. The underlying idea is that events occur preferentially (or on the opposite unfavourably) when they are located simultaneously inside two half-zones of both cycles. This hypothesis was demonstrated for instance by \citet{klein1976} who showed that there is a significant tendency for earthquakes in selected regions of the world to occur at times at which the tidal stresses are so aligned as to provide maximum enhancement of the stresses associated with plate tectonics. For the purpose, we represent a bi-axial plot of the two phase distributions in which we divide the diagram into four equal quadrants. As shown in Figure \ref{fig:phasevsphase_1}, for the case of precession phase versus the semi-diurnal phase (Figure \ref{fig:phasevsphase_1a}), there is a neat concentration in the upper left zone for earthquakes with M $\geq 6$. 
On the contrary, only one earthquake is found in the upper right zone. Note that such a probability $P$ to find only one event in one quadrant in the case of an even distribution is: 
\begin{equation}\label{eq:prob}
P = N \times {\frac{1}{4}} \times\left(\frac{3}{4}\right)^{N-1}   
\end{equation}
where $N$ is the total number of earthquakes. By applying this formula with $N = 32$ in our case we find $P = 1.0715 \times 10^{-3} \approx 1/933$. This very small value leads in favour of a bi-modal anti-correlation characterized by a very high probability that no event occurs when both phases (for solar-day and precession cycles) lie in the interval [0 - 0.5]. 

We remark that this trend also appears, but less neatly, when we plot the phase of the semi-lunar day vs. the phase of the fortnightly component (Figure \ref{fig:phasevsphase_1b}), with only four earthquakes in the upper right zone. Meanwhile, other bi-modal phase vs. phase configurations do not follow this trend and the earthquakes are more equivalently distributed. 

%%%%%%%%%%%%%%%%%%%%%%%%%%%%%%%%%%%%%
\section{$\chi^2$ test of correlation} \label{correlation}
%%%%%%%%%%%%%%%%%%%%%%%%%%%%%%%%%%%%%

In this section we apply the $\chi^2$ statistical test presented in Eq. \ref{eq:chi} to measure the presence of clustering in a quantitative way. For that purpose, we consider three kinds of data for estimating the $\chi^2$ value. First, we consider all the data starting from 1800 and restricting only for the 32 earthquakes with M $\geq 6$.  Second, we consider relatively accurate data starting from 1933 and we still limit our set to the 15 earthquakes with M $\geq 6$. Third, we compute the $\chi^2$ by considering the weaker earthquakes from 1933 with magnitude $5.3 \leq M < 6$. This choice is made necessary because some data before 1933 is not always accurate and presents a lack of homogeneity \citep{kilstonknopoff1983}.

\begin{table*}
\small
\caption{Value of the $\chi^2$ parameter by considering  different time range as well as different magnitude threshold.}
\begin{center}
{\renewcommand{\arraystretch}{1.1}

\begin{tabular}{lrcccc}
\hline
\hline
Cycle  & Period & 1800 - 2020 & 1933 - 2020 & 1933 - 2020    \\
       &        & $M \geq 6.0$& $M \geq 6.0$& $5.3 \leq M < 6.0$      \\
\hline
Precession      & 18.613 y.&       1.92        & 3.03        & 0.03         \\
Semi-diurnal  & 12 h.  &       2.26        & 7.62        & 0.17         \\
Semi-lunar day  & 12.421 h.&       0.15        & 1.22        & 0.01         \\
Fortnightly     & 14.765 d. &       0.42        & 2.16        & 2.24         \\
Annual          & 1 y.    &       0.03        & 0.01        & 3.93         \\
\hline
No. of events    &    & 32 & 15 & 36 \\
\hline
\hline
\end{tabular}}

\label{tab:chi2_earthquake}
\end{center}
\end{table*}

\begin{table*}
\small
\caption{Comparison of $\chi^2$ parameter from this study for the interval [1933-2020] with the one obtained by \citet{kilstonknopoff1983} (K \& K) for the 
interval [1933-1980].}

\begin{center}
{\renewcommand{\arraystretch}{1.1}

\begin{tabular}{lrccccc}
\hline
\hline
Cycle           & Period      &\multicolumn{2}{c}{$M \geq 6.0$} && \multicolumn{2}{c}{$5.3 \leq M < 6.0$} \\
                &             & This study & K \& K && This study & K \& K\\
\hline
Precession      & 18.613 y.   & 3.03       & 5.58 && 0.03 & 0.28 \\
Semi-diurnal  & 12 h.       & 7.62       & 5.20 && 0.17 & 0.49 \\
Semi-lunar day  & 12.421 h.   & 1.22       & 0.35 && 0.01 & 0.03 \\
Fortnightly     & 14.765 d.   & 2.16       & 8.49 && 2.24 & 1.23 \\
Annual          & 1 y.        & 0.01       & 0.05 && 3.93 & 0.46 \\
\hline
No. of events & & 15 & 9 & & 36 & 22 \\
\hline
\hline
\end{tabular}}

\label{tab:chi2_earthquake_comp}
\end{center}
\end{table*}

The results of our $\chi^2$ tests are displayed in Table \ref{tab:chi2_earthquake}. As mentioned in Sect. 2, in the same way as K \& K, we accept the null hypothesis for an homogeneous Poisson process at the 95 \% confidence level if $\chi^2 < 3.84 $, and at the 90 \% confidence level if $\chi^2 < 2.71$. For the data starting from 1800, there is no significant correlation between any of the five cycles considered and the earthquake events. However, if we limit our investigations to the data from 1933, with homogeneous and accurate values of the magnitude, we note both a significant correlation of strong earthquakes with
precession and a very big correlation with the semi-diurnal component ($\chi^2 = 7.62$). On the contrary, the $\chi^2$ value is close to zero when considering the weaker earthquakes dataset both for precession and semi-diurnal components, which tends to signify a lack of correlation. 

Surprisingly, the semi-lunar day component leads to small $\chi^2$ values whatever be the configuration, though the corresponding tidal amplitude is significantly larger than the semi-diurnal one. Note that the same paradoxical conclusion was already mentioned by K \& K who initially expected that a correlation should appear with the semi-lunar day cycle. Their argument is that the lunar tide is considerably stronger than the solar tide, and if a correlation is found with the semi-diurnal component, an equally strong correlation should be found with the Moon rise / set periodicity. The reason of an opposite result might be caused by the chosen period: indeed the lunar day is not constant, for we adopted a mean value.  So the level of correlation found should strongly depend on the adopted value. This problem is fully explored with more details in Section \ref{sensitivity}. The same remarks are available for the fortnightly cycle whatever be the configuration. Concerning the annual component, there is no correlation when considering the strong earthquakes dataset. Meanwhile, taking the earthquakes between 5.3 and 6.0 leads to a possible correlation ($\chi^2 = 3.93$).

Table \ref{tab:chi2_earthquake_comp} shows the comparison of our results with those of K \& K whose dataset did not take into account the earthquakes since 1983.  Even if the $\chi^2$  values are something different for precession and semi-diurnal components, they are in both cases above the threshold which leads to the conclusion of the existence of a correlation for the strong earthquakes.  On the opposite the lack of correlation with the semi-lunar day component is observed in both studies whatever be the dataset considered.

At last, both studies disagree on the fortnightly and annual components. For the first one, we find no significant correlation ($\chi^2 = 2.16$) whereas K \& K found a high correlation for strong earthquakes ($\chi^2 = 8.49$). This could be due to the methodology used : whereas we adopted a mean value of the period of the fortnightly cycle, K \& K referred the occurrence of the earthquakes to the times of the extrema, rather than to perfectly periodic occurrence. Concerning the annual cycle, we found a correlation with weak earthquakes ($\chi^2 = 3.93$) which was not reckoned by K \& K. This strongly raises the question of the variability of the results according to the choice of the time interval chosen for the dataset. 

%%%%%%%%%%%%%%%%%%%%%%%%%%%%%%%%%%%%%
\section{Sensitivity test} \label{sensitivity}
%%%%%%%%%%%%%%%%%%%%%%%%%%%%%%%%%%%%%

Following Eq. \ref{eq:chi}, it is clear that the $\chi^2$ value depends on the period chosen for the corresponding cycle. Indeed, for the lunar day and the fortnightly cycles, the period between two tidal maxima (or minima) varies in a significant way, whereas we chose their sole mean value to estimate $\chi^2$ in the previous section. Therefore, here we test the sensitivity of the $\chi^2$ value to the period of the cycle considered by re-estimating this parameter through the interval between its minimum and maximum values of the period. We concentrate our study on the two cycles above. Notice that the lunar month has an average period of about 29.53 days and varies approximately within the interval $\pm$ 7h. So, the minimum value of the fortnightly period is 14.59 d. whereas the maximum one is 14.91 d. Meanwhile, the variation of semi-lunar day is about $\pm$ 9 mn. Hence, the minimum and maximum periods are respectively 12.27 h. and 12.57 h. Here we restricted the calculations by using the earthquakes from 1933 and $\geq 6$ magnitude, for the sake of homogeneity. 

\begin{figure*}
	\centering
	\begin{subfigure}[t]{0.49\textwidth}
		\caption{\textbf{$\chi^2$ semi-lunar day cycle}}
		\includegraphics[width=1\textwidth]{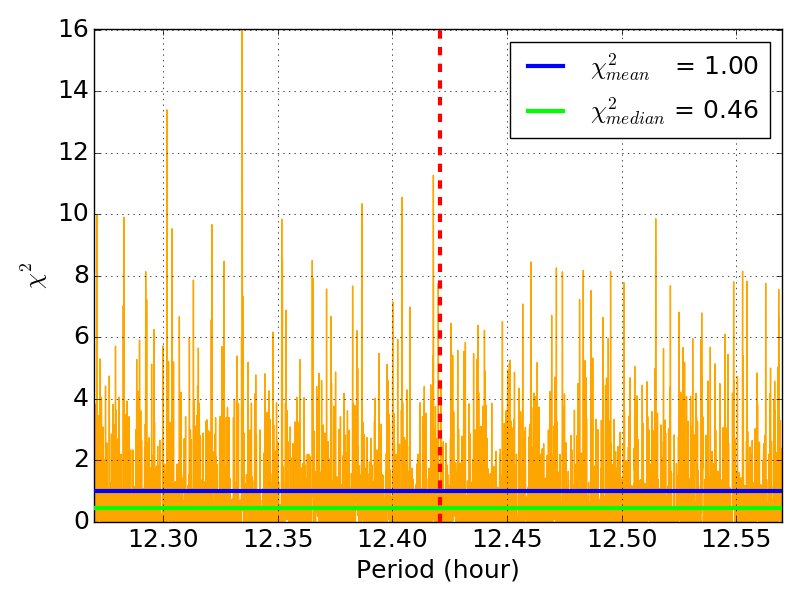}
		\label{fig:event_1233a}
	\end{subfigure}
	\begin{subfigure}[t]{0.49\textwidth}
	\caption{\textbf{"12.33 hours" phase}}
		\includegraphics[width=1\textwidth]{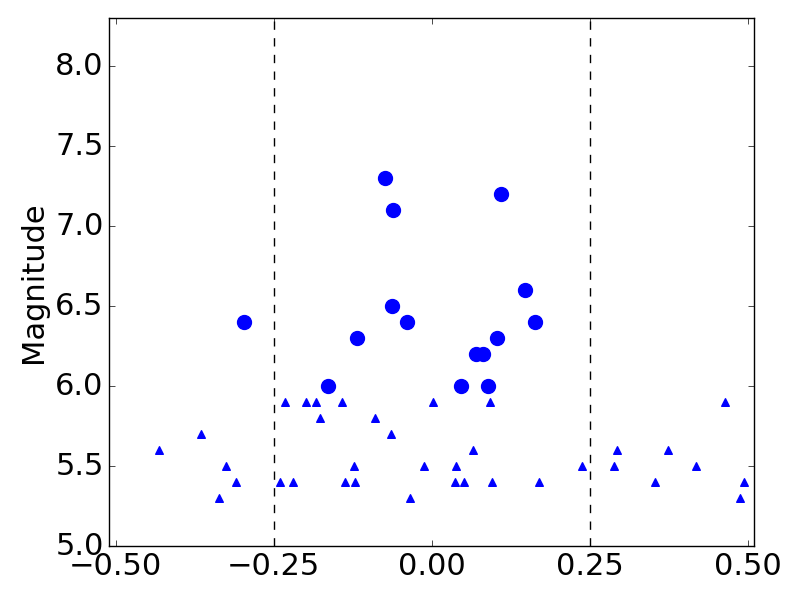}
		\label{fig:event_1233b}
	\end{subfigure}
	\par\bigskip
		\centering
	\begin{subfigure}[t]{0.49\textwidth}
			\caption{\textbf{$\chi^2$ fortnightly cycle}}
		\includegraphics[width=1\textwidth]{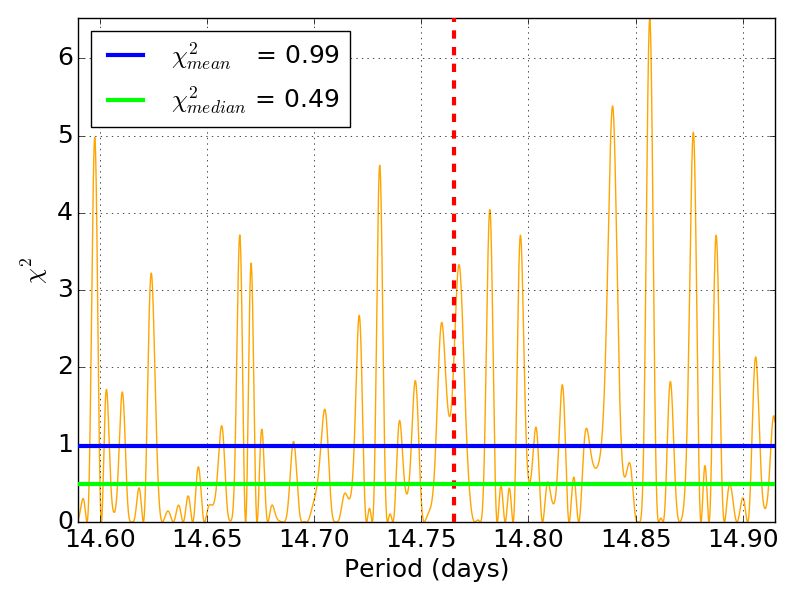}
		\label{fig:event_1233c}
	\end{subfigure}
	\begin{subfigure}[t]{0.49\textwidth}
			\caption{\textbf{"14.86 days" phase}}
		\includegraphics[width=1\textwidth]{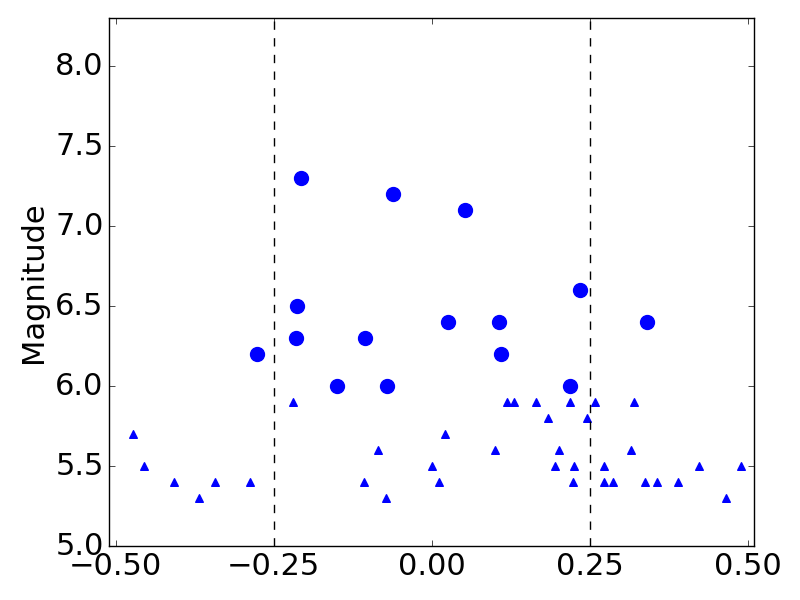}
		\label{fig:event_1233d}
	\end{subfigure}
	\caption{\textit{Left}: $\chi^2$ estimation between the minimum and maximum values of the semi-lunar day period (a) and fortnightly (c). Red dashed line represents the nominal value that we used in Section \ref{correlation}. \textit{Right}: phase distribution with respect to the phase of the semi-lunar day with period 12.33 hours (b) and the phase of fortnightly with period 14.86 days (d). Here \textcolor{blue}{$\bullet$} represents the earthquake with M $\geq 6$ and \textcolor{blue}{$\blacktriangle$} represent the earthquake with M $< 6$.}
\label{fig:event_1233}
\end{figure*}

As shown in Figure \ref{fig:event_1233a}, the value of $\chi^2$ concerning the semi-lunar day correlation varies between 0 (period = 12.39 hours) and 16 (period = 12.33 hours), thus signifying a high sensitivity with respect to the period considered. This statement is emphasized by the presence of numerous peaks in the interval.
  
To illustrate the high level of correlation found for period 12.33 h. we show in Figure \ref{fig:event_1233b} the corresponding phase distribution of the earthquakes. For the 15 earthquakes with  M $\geq 6$, the phase distribution is much more centralized in the middle half of the phase. Only one earthquake is found outside. By adopting the same formula as Eq. \ref{eq:prob} with N=15, we obtain a probability of P = 6.68 \%  that such a distribution is obtained by considering a null hypothesis. However, we remark that such a clustering besides being still apparent, is less obvious for earthquakes with M $< 6$.

Following a similar procedure, Fig. \ref{fig:event_1233c} depicts the variation of $\chi^2$ in the range of possibilities of the value of the fortnightly period. Here $\chi^2$ varies from almost 0 (for a 14.81 d. period) to a rather limited value of 6.5 (for 14.86 d.). Figure \ref{fig:event_1233d} shows the phase distribution with the period = 14.86 d. corresponding to the maximum. In a similar way to Fig. \ref{fig:event_1233b}, the earthquake phase is located in the middle half of the figure for the earthquakes with M $\geq 6$. Accordingly we suggest being more cautious for interpreting our results related to $\chi^2$ from semi-lunar day and fortnightly periodicities. A more rigorous method needs to be applied to study the correlation between these two cycles and earthquakes events, in particular by computing the phase by interpolation with respect to two successive extrema of the cycle.
 
\begin{figure}
\centering
\includegraphics[width=0.7\hsize]{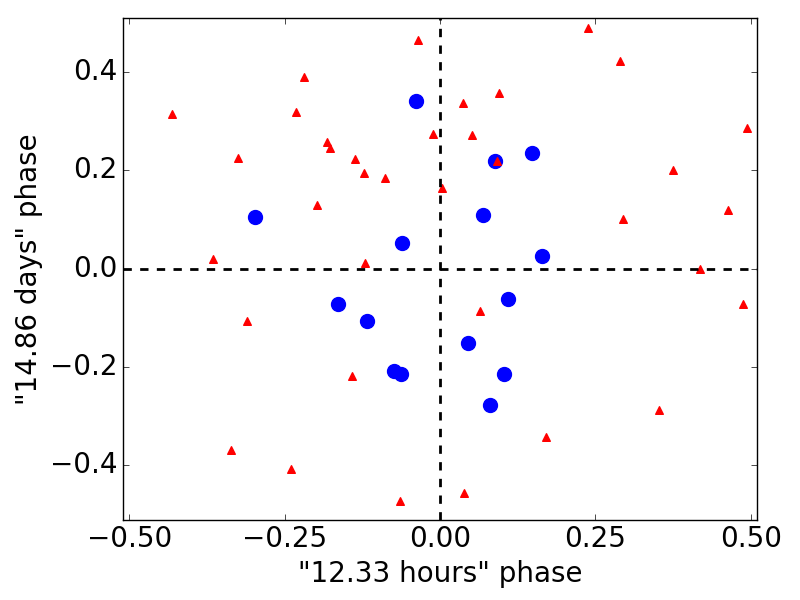}
\caption{Earthquake phase distribution in fortnightly with period = 14.86 d. vs. the phase distribution in semi-lunar day with period 12.33 h.
Earthquakes with M $\geq 6$ are represented in \textcolor{blue}{$\bullet$},those with M $ < 6$ M are in \textcolor{red}{$\blacktriangle$}.}
\label{fig:phasevsmag_2}
\end{figure}

Furthermore, it is worthwhile to analyze the phase vs. phase relation between the maximum $\chi^2$ period of semi-lunar day (period = 12.33 hr.) and of the fortnightly period (14.86 d). As shown in Fig. \ref{fig:phasevsmag_2}, the earthquakes with M $\geq 6$  are almost evenly distributed among four different quadrants. However, they are significantly concentrated in the centre of the plot. It likely signifies that the probability that an earthquake occurs becomes higher when the two events are at their peak (zero phase). Besides these two cycles, there is no meaning to carry out such a $\chi^2$ sensitivity test for precession, semi-diurnal and annual cycles, for they do not exhibit significant period variability. 
 
%%%%%%%%%%%%%%%%%%%%%%%%%%%%%%%%%%%%%
\section{Conclusion} \label{conclusions}
%%%%%%%%%%%%%%%%%%%%%%%%%%%%%%%%%%%%%

In this study, we re-conducted a work carried out nearly four decades ago by K \& K to investigate the possibilities of correlation between some lunisolar cycles and earthquake events that occurred in the Southern California region. We beneficiate of a considerably larger and more accurate data than in this last study. First, we analysed the phase distribution of the earthquakes with respect to the leading lunisolar cyclic tidal components. The corresponding periods are the semi-diurnal one (12 h.), the semi-lunar day (12.42 d) the fortnightly period(14.77 d), the year (365.25 d) and the period of the precession of the nodes of the orbit of the Moon (18.613 y.). We remarked that apparent clustering of earthquakes with M $\geq 6$ occurs near the zero phase in the diagrams related to the precession and semi-diurnal cycles, thus confirming a statement already established by K \& K. Moreover, we analysed the possible link between these two cycles. We found that there is a high probability that no event occurs when both phases (for semi-diurnal and precession cycles) lie in the interval of [0, 0.5], with only a 0.11 \% probability of the studied distribution to happen when adopting the hypothesis of a lack of correlation. 

In order to confirm the tendencies found above, we applied the same $\chi^2$ statistical test as that established by K \& K to validate the presence of clustering in a more rigorous way. We found that when limiting our investigations to events between 1933 and 2020 for which the quality of data is quite homogeneous and particularly accurate, strong earthquakes with M $\geq 6$ present a high level of correlation with semi-diurnal and precession phase, which clearly confirms K \& K conclusions. In addition, we suspect also the presence of a significant correlation between weaker earthquakes ($5.3 \leq M < 6.0$) and the annual cycle, which was not pointed out by K \& K. At last there is apparently no correlation of earthquakes with the fortnightly and semi-lunar day tidal cycles, despite their strong respective tidal amplitudes. To measure the robustness of our result, we re-estimated $\chi^2$ by varying the period of the fortnightly and semi-lunar day cycles between their minimum and maximum values. Thus we found that the $\chi^2$ results for both cycles are not robust. In other words, the value of $\chi^2$ can assert a total absence or a high level of correlation according to the value of the period selected. Therefore a more rigorous statistical method is needed to study these two cycles, which should consider phase shifts due to the variability of the periods considered. These phase shifts, if not taken into account, would lead to misidentification of the times of peaks of tidal components, as already explained by K \& K. Such a sensitivity test is not necessary for the annual, diurnal (12h.), and precession components, which have a well established and not varying period. 

An important output from the present study is that we confirm the main results found by K \& K a little less than four decades ago concerning the interaction of luni-solar extensional Earth tides in a part of the Earth where strike-slip faulting occurs with a well-defined geographical orientation. It should be interesting to extrapolate our positive results concerning a tides-earthquakes correlation for the semi-diurnal and precession cycles to anticipate the time of the next episode of high potential for the occurence of large earthquakes in southern California. 

%REFERENCES
\bibliographystyle{roaj}
\bibliography{nurulhuda_and_souchay_2021_roaj}
\makeatletter
\def\@biblabel#1{}
\makeatother
%-----------------------------------------------------------------------
\received{\it 2 December 2020}
\end{document}